# Learning Each Other's Ropes:
# Negotiating interdisciplinary authenticity

Edward F. Redish* and Todd J. Cooke**
*Dept. of Physics, U. of Maryland, College Park MD 20742
**Dept. of Cell Biology & Molecular Genetics, U. of Maryland, College Park MD 20742

**Abstract**

A common feature of the recent calls for reform of the undergraduate biology curriculum has been for better coordination between biology and the courses from the allied disciplines of mathematics, chemistry, and physics. Physics has lagged math and chemistry in creating new biologically oriented curricula, though much activity is now taking place and significant progress is being made. In this article we consider a case study: a multi-year conversation between a physicist interested in adapting his physics course for biologists (Redish) and a biologist interested in including more physics in his biology course (Cooke). These extended discussions have led us both to a deeper understanding of each other's discipline and to significant changes in the way we each think about and present our classes. We discuss two examples in detail: the creation of a physics problem for a biology class on fluid flow, and the creation of a biologically authentic physics problem on scaling and dimensional analysis. In each case, we see differences in how the two disciplines frame and see value in the tasks. We conclude with some generalizations about how biology and physics look at the world differently that help us navigate the minefield of counterproductive stereotypical responses.

## I. Preamble and Motivation: The call

As this special issue of CBE indicates, there is considerable interest and activity in the transformation of undergraduate biology education. Calls from the community of research biologists (National Research Council, 2003; 2009; AAAS, 2011) and health-care professionals (AAMC/HHMI, 2009) have been followed by reform, both in biology courses themselves, and in the science courses supporting them. At the University of Maryland, new courses have been developed in Organismal Biology [BSCI 207], General Chemistry [CHEM 131-132, 271-272], and Math for Biologists [MATH 130-131]. Physics as a national community has been a bit slower in developing physics classes explicitly meant to serve all biologists, but there is now a strong and growing interest in the physics education community.[1]

At Maryland, the two authors, a biologist (Cooke) and a physicist (Redish), have been interacting on the topics of physics in biology classes and physics class for biologists since 2005. These interactions have led us to reform both our own biology and physics classes and, more recently, to participate in a larger project creating a physics course specifically designed for life and health science majors as part of the National Experiment in Undergraduate Science Education (NEXUS) supported by the Howard Hughes Medical Institute (HHMI).

Our interactions both with each other and with other faculty in biology, physics, chemistry, and mathematics through the NEXUS project have taught us the importance of understanding the differences between our disciplinary perspectives – "learning each other's ropes." In this paper we use our experience as a case study. We provide detailed specific examples that illustrate the disciplinary differences that surprised us, and we develop broad heuristics that can help our disciplines learn how to go forward with mutual understanding and respect.

## II. The Case Study

One thing that we have learned in our years of interaction is that interdisciplinary reform of physics and biology classes will require significantly more than occasional discussions or sending out surveys asking faculty in each discipline, "What do you think is important to include?" Each discipline brings its own distinct perceptions as to what content is important for introductory classes, what epistemological orientation to bring to those classes, and what competencies are appropriate to develop in what order. Often these perceptions are tacit. Only through frequent conversations can these hidden assumptions be brought to light.

In the years we have known each other, we have participated in more than 500 hours of intense conversation on the issues of how to bring together biology and physics in introductory classes. While our experience is not unique in being an extended collaboration between a physicist and a biologist,[2] our collaboration is somewhat singular in that one of us (Redish) has spent a significant fraction of his research career (from 1992 to the present) in discipline-based education research and has significant access to and interaction with educational specialists. This allowed us to quickly construct a biology education research group (UMd-BERG[3]) and merge it with an existing physics education research group (UMd-PERG[4]) that has a long track record and access to first-class graduate students and postdocs. We therefore think it worthwhile to tell our story in detail and document some of our experiences and some of what we have learned.

**Close encounter of the third kind**

Despite both having been faculty members at the University of Maryland for decades, our protagonists first met in the fall of 2005. Each had applied to join the University's newly formed *Academy for Excellence in Teaching and Learning*, a group of senior faculty who were interested in improving education on campus.

At the beginning of the fall term, about a dozen faculty gathered at the group's first meeting of the year. As we went around the table introducing ourselves, an Emeritus Physics Professor involved with developing classes and support for in-service high-school science teachers introduced himself. Our co-author biologist followed, his introduction including a rant on physicists in general: "You physicists might do a great job teaching physics, but my biology students can't use it in our biology courses." When the introductions came around to our co-author physicist, he reported, "I have been working on reforming the physics for biologists class for five years and here are the names of the biologists I have talked to."

After the meeting we immediately went to speak to each other. Our biologist said to our physicist, "I didn't know anybody like you [i.e., a physicist with a real interest in serving the needs of biology] existed." Our physicist, delighted to find a biologist actually interested in using physics in his biology course, responded, "Let's have lunch." We began to meet regularly and within a few years were getting together every week to discuss issues in biology, physics, their interaction – and how to reform our pedagogy to produce the best results.

We quickly learned that each of us had oversimplified views of the relation between teaching physics and teaching biology. The failure of students in the biology class to know much physics could have been expected since physics had not been made a prerequisite for that course and most of the students had not previously taken college physics. On the other hand, even if they had taken physics, the traditionally offered content would not have helped them much for the tasks our biologist was interested in.

Our physicist was reforming his course with biologists in mind. (It's the "epistemologized physics class" described below.) Although his reforms had been successful along many dimensions (Redish & Hammer, 2009), the reforms did not include meeting the explicit content needs of the biologists and conveying how to use physics to address biological problems.

Over the years since our meeting, each of us has wound up modifying our innovations to reflect what we have learned from each other. Three years ago, we began the NEXUS project: creating a *Physics for the Life Sciences* class "from scratch" through extensive negotiations between physicists and biologists. We brought in dozens of additional scientists – biologists, physicists, chemists, and mathematicians – into our discussions. In this paper we summarize what we have learned from hundreds of hours of these interdisciplinary conversations.

To give a view of our starting points, in the remainder of section II we briefly describe the reforms we each independently carried out to try to include physics in a biology class and to create a physics class appropriate for biologists. In section III we offer two detailed examples showing how our interactions resulted in changes in each of those classes. In section IV we discuss a detailed example from the NEXUS project showing how the interdisciplinary perspective informs our decisions and approaches. In section V we generalize, creating heuristics about the differences in the disciplinary approaches. Our conclusions are given in section VI.

**Independent reform of a biology class to include physics**

At many colleges and universities, the introductory biology sequence is divided into three major courses:

1) molecular and cell biology, including biochemistry and genetics,
2) organismal biology, including the diversity and functions of organisms, and
3) ecology and evolutionary biology.

When several biology departments at the University of Maryland created a common introductory biology sequence in the mid-1980s, these departments reached the consensus that this sequence ought to consist of two courses to teach the fundamental principles of molecular biology and ecology, but left unresolved the vexing question of how to teach organismal biology. Conventionally, organismal biology is taught as "a forced march through the phyla" that consists of separate units on the distinguishing characteristics of each major group of organisms, which are followed by separate units on animal and plant physiology.

The solution adopted at Maryland was to require each undergraduate biology major to take one specialized course focusing on the diversity and/or function of a single group of organisms, such as the microbes or the animals. Given that many processes having great biological and medical significance involve the interactions of different organisms, this solution was unsatisfactory in retrospect, but it persisted for almost 20 years.

In 2004, our biologist was appointed to chair a committee of biology faculty who were teaching these specialized diversity and physiology courses. The charge given to that committee was to identify the common principles governing the biology of all organisms that might, in turn, serve as the basis for a third course in the introductory biology sequence. Eventually, this committee created the syllabus for a *principles-based organismal biology course* designated as BSCI 207.

BSCI 207 focuses on the physical, chemical, genomic, and evolutionary principles that account for the unity and diversity of all life. For example, one principle emphasized in BSCI 207 is that all non-life and life are governed by universal mathematical, physical, and chemical principles. These principles include thermodynamics, transport processes (sometimes called gradient-driven flows: diffusion, fluid flow, electricity, and heat transfer), oxidation-reduction, scaling, material properties, and mechanics. A second principle involves deep molecular homology: all living organisms are descended from a common ancestor (or common ancestral community). Thus, organisms share a common genomic tool kit encoding for homologous molecules that regulate the mo-lecular activities of life. BSCI 207 is structured around these and other principles, which are illustrated and/or explicated by their expression in all major groups of organisms.

To the faculty teaching BSCI 207, this emphasis on organizing principles felt like a much more rewarding and effective approach toward teaching organismal biology than the traditional "forced march." Our biologist started receiving recognition for teaching innovation from different campus groups.

But then he met our physicist, who had the temerity to ask, "How do you know if the BSCI 207 students are actually using these principles to organize their knowledge and if they can apply them in new situations?" After initially dismissing these questions, our biologist began to interview BSCI 207 students, who reported that they viewed these principles as "just more stuff to memorize for the tests."

What started off as our biologist criticizing the efforts of the physicists trying to teach biology students had come full circle – our biologist had as much to learn from our physicist about how to teach physics to the biology students as vice versa. The resulting conversations became the basis for an NSF grant to reform the pedagogy in the BSCI 207 class.[5] This class, which we refer to as the *principle-based organismal biology class*, forms the context for the biologist's side of the first level reforms arising from our interactions in this case study.

**Independent reform of a physics class for biologists**

Our physicist has been interested in the question of how to teach physics for biologists since the first time he was assigned to teach the class as a young faculty member in 1975-76. The following summer, he was sent to represent his department at a conference at MIT on the subject of "Teaching Physics for Related Sciences and Professions" (French & Jossem, 1976) where he attended the sessions on Physics for the Life Sciences and Physics for Biomedical Students. There he first had the opportunity to ask his biology colleagues and med school faculty, "What do you want me to teach?" He almost always got the response, "Teach them to think like physicists."

But as a physicist, he didn't know what that meant. He doesn't "think like a physicist," he just thinks. And the orientation he brings to his thinking about the physical world comes from years of education and research experience. Can any of this be taught in two semesters to biology majors and pre-meds? For many years, when given the opportunity to teach algebra-based physics (the traditional introductory physics course that includes biologists and pre-meds), he included "biological examples" wherever possible, but otherwise taught the standard class.

Between 2000 and 2005, our physicist and his colleagues in the UMd-PERG undertook a major reform of the traditional algebra-based physics class with an emphasis on making it more appropriate for life science students. This involved a combination of basic research and materials development supported by a series of NSF grants.[6]

The UMdPERG's previous research (Redish, Saul, & Steinberg, 1998) had demonstrated that students often brought into introductory physics *epistemological misconceptions*: their expectations as to what kind of knowledge they were learning and what they had to do to learn it. These were often poorly aligned with instructors' goals and expectations. For example, students often assumed that they were learning lots of independent factoids ("flash cards" or "equation sheets") rather than a coherent, principle-based reasoning method. They often assumed that sense-making and strong conceptual knowledge were irrelevant and that memorizing equations for calculational purposes was all they needed to do.

This reform of algebra-based physics emphasized shifting student *epistemologies*, attempting to shift their view of learning from pieces to coherence, from equations to concepts, from externally-transferred knowledge to internally-generated reasoning, and from treating their everyday think-ing as irrelevant to reconciling their experience with the physics they were learning. To do this, the project "epistemologized" best-practices research-based instructional methods, but did not significantly modify the content to accommodate the needs of life science students, in part because the course served other populations as well. The reformed course showed strong gains on both standardized conceptual and attitudinal measures. (Redish & Hammer, 2009) This course, which we refer to as *the epistemologized physics course*, forms the physicist's side of the first level reforms arising from our interactions in this case study.

## III. Detailed Examples: Level 1

Despite having met and interacted in 2005-06, our protagonists each had a sabbatical year so serious interactions were only begun in 2007-08. At that point, each of us had the major course reforms describe above under our belts. But our subsequent interactions led us each to make changes in those courses. Here are two detailed examples.

**Modifying the existing physics class: Shifting the content (The H-P Equation)**

Our first detailed example comes in the context of our physicist's epistemologized physics class. Traditional introductory physics includes a very limited discussion of fluids – mostly statics including the increase of pressure with depth, Archimedes' principle and buoyancy, with perhaps a brief discussion of fluid flow including Bernoulli's principle (with the famous demonstration of a beach ball levitated by an air stream).

One of our first discussions about the physics class brought a complaint from our biologist that it "didn't cover the topics biologists need – for example fluids." In our biologist's principle-based organismal biology class, he emphasizes the importance of gradient driven flows, such as the flow of fluids described by the Hagen-Poiseuille (H-P) equation, which is critical for understanding the long-distance transport of fluids in large animals and plants.

In BSCI 207, the parameters of flow, pressure, and resistance help students understand the fundamental mechanisms and functional differences in fluid flows in an animal vs. a plant; for example, in an acacia tree and in the giraffe that eats its leaves. Of particular significance is that the resistance of a segment of pipe is inversely proportional to the fourth power of the radius; in contrast to the resistance of an electrical resistor, which is inversely proportional to the square of the radius. This fourth-power relationship is crucial for understanding why arteriosclerosis, or the depositing of plaque inside blood vessels, has such deadly consequences for humans.

Our biologist focuses on this particular functional dependence as it has powerful implications for the biology. Students in BSCI 207 often cite this fourth-power relationship when they are asked on the final exam to support or refute the statement that physics governs the functioning of organisms.

Moreover, they seem satisfied with a phenomenological explanation, at least in the context of the biology class, because they have never inquired about its derivation. Nor does our biologist care, in part because he knows that Poiseuille was trained as an experimental physiologist who discovered the four-power relationship from his experiments working first with animal blood vessels and then with glass tubes. (Sutera & Skalak, 1993)

Our physicist was not satisfied with this phenomenological approach and wanted to develop a better understanding of *why* there was a difference between Ohm's law and the H-P equation. A detailed analysis of how the two protagonists think about the equation helps us to see the difference between what satisfies our biologist and what our physicist wants. This goes to the heart of the differences between what each values. We begin with the physicist's analysis of the H-P equation. The biologist's take on it is discussed in the following subsection.

Consider a pipe with a uniform circular cross section containing a fluid flowing at a constant velocity. Let's work with a toy model (*i.e.*, a drastic oversimplification) that assumes the fluid flows uniformly in the pipe – each bit of the fluid moving with the same velocity.[7]

Consider a thin cylinder of fluid that is moving down the pipe (colored blue in figure 1 below). Since the fluid is moving at a constant velocity, the forces on each bit of the fluid are balanced. Because we assume that there is some frictional or drag force proportional to the velocity that tends to slow the fluid down, this must be balanced by a force that tends to speed the fluid up. This tells us that the pressure must drop in the direction of the flow, so the force on each bit of fluid from upstream is larger than the force from downstream. When the fluid is moving at a constant velocity, according to Newton's laws of motion, this difference of the pressure forces must exactly balance the drag.

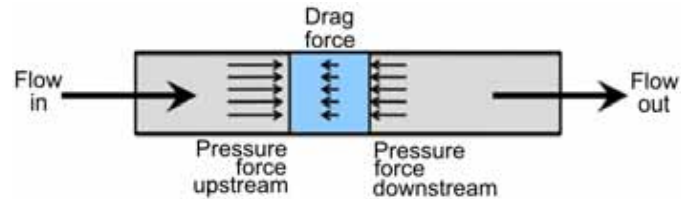

*Fig. 1: The toy model for explaining the source of the fourth power of the radius in the HP equation.*

Let the cross sectional area of the pipe be $A$ (= $\pi R^2$), the up- and down-stream pressures $P_U$ and $P_D$, and the velocity of the fluid $v$. Taking a simple model of the drag force as proportional to the velocity and the length, $L$, of the cylinder (consistent with what we know about viscosity) we get that the balance of forces looks like:

$$F_{pressure} = P_U A - P_D A \qquad F_{drag} = bLv$$
$$(P_U - P_D)A = bLv$$

where $b$ is some coefficient proportional to the viscosity. (The details don't matter.) Writing the pressure difference as $\Delta P$ and noting that the volume current flow, $Q$, is given by the cross-sectional area of the pipe times the speed of the flow:

$$Q = Av$$

so we get the result

$$(\Delta P)A = bL\left(\frac{Q}{A}\right)$$
$$\Delta P = \left(\frac{bL}{A^2}\right)Q = ZQ$$

This is the H-P equation with the resistance constant $bL/A^2$ defined as $Z$.

From the foregoing analysis it becomes clear *why* the H-P equation has a stronger dependence on the radius than does the resistance in Ohm's law, which is only proportional to $1/A$. The difference comes from what is physically (and biologically!) significant in the two cases. In the electric current case it is the force per unit charge that matters, leading to a voltage difference. In the fluid case it is the force per unit area that matters, leading to a pressure difference. What matters is not just the fact that the voltage and the pressure are what is easy to measure; what matters is the fact that in a biological system it is the pressure that is carried through the system and has implications.

Our discussions about the H-P equation led us to negotiate a pair of matched problems that we have used in both the physics and biology classes.[8] The version for biology is shown in figure 2.

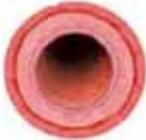

*Fig.2 : The H-P problem for the biology class.* (Image source: http://www.nlm.nih.gov/medlineplus/ency/imagepages/18020.htm)

**Modifying the existing biology class: Shifting the pedagogy**

An important shift in the pedagogy occurred in the context of our biologist's principle-based organismal biology class. Our biologist realized from his interactions with our physicist that conventional lectures are ineffective for teaching the general principles that are useful for explicating the evolution and functioning of organisms. Instead, it was necessary to modify the active-learning tutorial approach often used in physics classes to help students master qualitative reasoning. (McDermott et al. 1998; McDermott & Shaffer, 1992; Redish, 2003 pp. 146-156.)

As a result of our discussion, students in BSCI 207 are now assigned to permanent small groups within the large classroom. A significant number of the class periods are devoted to what are called Group Active Engagements (GAEs). Each GAE is designed to help the student groups discuss and then organize their knowledge into a conceptual, physical, and/or mathematical (*i.e.*, an equation) model of a particular principle. Then the groups are asked to apply this model toward a deeper understanding of biological phenomena during the remainder of the class period as well as during homework problem-solving sessions where they meet on their own outside of the class.

Although GAEs are also used to teach non-physical principles, e.g., origin of life, endosymbiosis, and life cycles, here we briefly describe the GAE for a physical principle here, namely, the Hagen-Poiseuille equation for fluid flows in large plants and animals – the same equation we just derived for the physics class.

In this GAE, the students are shown a picture of a giraffe straining to reach the underside of an acacia tree. It turns out that a giraffe eats more than 30 kg of acacia leaves every day, which serves as its major source of nutrients and water. The students are asked to argue in their groups about which organism has the more "powerful" pump. This question is purposefully ambiguous because we want the students to argue about how to describe the process of fluid flow. Frequently, the students within each group end up weighing two possible answers: the acacia ("because it is taller than a giraffe and must work harder against gravity" or the giraffe ("because the blood spurts out when you cut a giraffe, but not when you cut a tree").

The argument for the acacia is essentially emphasizing the pressure required to move bodily fluids, whereas the argument for the giraffe focuses on their flow rate. The groups are then asked to reconcile those two answers, which they will eventually recognize as involving a third parameter, the resistance, following subtle or sometimes direct hints about the parallels of fluid flow to Ohm's Law. Thus, they have developed a skeletal version of the Hagen-Poiseuille equation, or[9]

$$\frac{V}{t} = \frac{\Delta P}{R}$$

where $V/t$, $\Delta P$, and $R$ are the flow rate, pressure difference, and resistance, respectively. Plants are further characterized as having evolved vascular systems capable of generating high pressures (e.g., up to 7000 kPa for water transport in trees) in the presence of much higher resistances. The hearts of animal circulatory systems generate output pressures several orders of magnitude lower (e.g., 120/80 mm Hg or 16/11 kPa in human systemic circulation). But animals have also evolved blood vessels having much lower resistances, thereby resulting in much higher flow rates.

However, that dependence on very low resistances for generating high flow rates makes animals extremely vulnerable to arteriosclerosis. This brought home to the students by having them do the problem in Fig. 2 involving the consequences of a 50% decrease in the radius of the open lumen of a coronary artery. The physics of the expanded version of H-P equation is non-negotiable here. The flow rate in such occluded arteries under constant blood pressure must decrease as a fourth-power function resulting in a 93.75% lower flow rate. The human body attempts to compensate for arteriosclerosis by increasing its blood pressure, but pressure is linearly related to flow rate so that compensating pressure in this example is calculated as being an impossible 16-fold higher pressure differential.

It is worth noting how the presentation of physical principles in BSCI 207 differs from the typical derivation in physics classes. Instead of showing the students how to derive the equation of interest from other equations, each GAE encourages the students to use their prior biological and everyday knowledge to identify, discuss, and understand the parameters before constructing the equation. Secondly, the GAEs do not typically use an idealized model for first illustrating the principle, but rather attempt to use simplified descriptions of real organisms to maintain an obvious connection between physics principles and biological phenomenon.

Lastly, little effort is devoted toward fitting the equation into the broader conceptual foundations of physics. Students see little value in our initial efforts to relate the structure of the H-P equation to the common structure of all equations for gradient-driven flows. Instead, the utility of each equation for understanding the fundamental mechanisms operating in various biological phenomena is kept front and center.

## IV. Detailed Examples: Level 2

**Starting from scratch (almost)**

*The challenge of the NEXUS Physics class*

In 2009, the Association of American Medical Colleges (AAMC) working with HHMI published *Scientific Foundations for Future Physicians* – a call for rethinking education for biologists and pre-medical students in the US to bring in more and better coordinated science – biology, math, chemistry, and physics – and to focus on scientific skills and competencies. The result was the HHMI-funded *Project NEXUS*: the National Experiment in Undergraduate Science Education, a 4-year, 4-university $1.8 M project. (Thompson et al., 2012)[10]

At the University of Maryland, College Park (UMCP) we have opened an interdisciplinary conversation to create a physics course designed to meet the needs of biologists and pre-health-care-professionals.[11] We've put together a team of nearly 40 professionals, including physicists, biologists, chemists, and education researchers both on and off campus. Over the past two years, we have held hundreds of hours of interdisciplinary conversations and negotiations among subgroups of this team. We quickly discovered that creating an interdisciplinary physics course that meshes with what is being taught in biology and chemistry and meets the needs of life science majors is not going to be simple.

### *Competing perceptions of physicists and biologists: Rethinking content*

We learned that there are significant cultural differences between biologists and physicists. Some biologists see most of the traditional introductory physics class as useless and irrelevant to biology. Some physicists consider that most of introductory physics is "privileged" – a coherent structure appropriate for anyone who needed physics, with little room for change.

Extended negotiations between the two groups led each to understand aspects of the others' viewpoints. Physicists realized that much of the traditional content was taught with two hidden assumptions: (1) that the class should look like a basic introduction for professional physicists; introducing, if briefly, all of the fundamental concepts that would be elaborated on later in a curriculum for majors; and (2) that complete chains of mathematical reasoning should be emphasized and phenomenology minimized. However, when we considered the content in the context of biology students, these assumptions seemed inappropriate.

First, it is clear that very few biology and pre-medical students will be completing a full undergraduate major in physics. It is therefore inappropriate to design a course for them using assumption (1).

Second, there are many topics to which biology and pre-medical students will be introduced in biology and chemistry classes that they find difficult and where a physics-style approach could possibly help. Excluding these topics because a first-principles mathematical derivation is not possible seems inappropriate.

Two examples that illustrate these assumptions from the design of a traditional physics class are circular motion and diffusion. Circular motion can be successfully derived mathematically from fundamental principles discussed in introductory physics (kinematics and Newton's laws) either with or without calculus. A complete mathematical discussion of diffusion requires partial differential equations and techniques not accessible to most of our target audience.

Circular motion is included and diffusion is omitted in essentially all introductory physics classes for biologists. But circular motion has extremely limited and only highly technical applications in biology and medicine: understanding the operation of instruments such as centrifuges, mass spectrometers, magnetic-resonance imagery, etc. On the other hand, an understanding of diffusion (and random motion in general) plays an essential role in students' development of an accurate picture of the functioning of a cell.

Our interdisciplinary discussions led us to some dramatic changes in our thinking about what a physics course for biologists should look like. We dropped or significantly reduced canonical topics that could be treated mathematically in full, such as circular motion, projectiles, and collisions, while adding topics that depended on a more phenomenological introduction, such diffusion, molecular interactions, and discrete (quantized) excited states.

### *Competing perceptions of physicists and biologists: Rethinking examples*

In our interdisciplinary discussions we also learned that biologists and physicists had dramatically different views of what makes a good biological example in physics.

There are dozens of physics textbooks "for the life sciences". Many of these mark some of their end-of-chapter problems for biological relevance. But when these are considered by biologists, essentially all are rejected as trivial or as having no biological interest. The typical physics problem for the life sciences looks something like using a pumpkin as a projectile ("punkin' chunkin'") and doing a standard range calculation. Another classic failed example is the problem that identifies the energy of the photon absorbed by chlorophyll and asks the student to calculate the frequency and the wavelength, expressing the wavelength in a variety of different units. In both examples, although an object involved in the problem is biological, what one learns from doing the problem yields no new biological insight.

We came to understand that what would be of value in a physics class is *biological authenticity* – examples where solving a physics problem in a biological context gives the student a deeper understanding of why the biological system behaves the way it does. (Watkins, *et al*., 2012) One such example that we deemed biologically authentic is the implication of artery narrowing on blood pressure discussed in our sections on the H-P equation. In a number of cases, we have gone through extensive negotiations, seeking both physical validity and biological authenticity. These examples give insight into what each disciplinary culture values. In the next two subsections of the paper, we discuss two of these negotiated examples in detail.

### *Chasing a pronghorn*

When seeking a problem with a biological context for a midterm exam on kinematics in his epistemologized physics class, our physicist was inspired by a sentence from a book on cognitive linguistics: "The pronghorn antelope in the Western Great Plains of the United States is one of the fastest animals on the planet. But it has outlived all its predators and now it runs where none pursues." (Fauconnier & Tuner, 2003) He constructed a problem in which a cheetah-like predator saw a running herd of pronghorn and decided to give chase. The cheetah can accelerate to a higher speed than pronghorns can maintain but can only hold that speed for a short time. The students were asked to calculate how far away the antelopes needed to be from the cheetah when the cheetah started its chase in order to escape their predator.

The cheetah-pronghorn problem has good physical authenticity, requiring the students to know the equations for constant velocity motion and constantly accelerated motion, and to be able to set up a multi-step mathematical problem from a physically described context.

Our biologist rejected this problem as biologically inauthentic. Our physicist, while looking up correct speeds and accelerations for the pronghorn and cheetah, had assumed (for the sake of the calculation he wanted the students to do) that the cheetah could continue to sprint at top speed for 30 seconds and then would drop down and be able to maintain a slower speed at steady pace. Our biologist, knowing more about cheetahs, explained that the cheetah's metabolic rate becomes so high during its acceleration that its temperature spikes to unsafe levels and it cannot maintain a steady run after a short chase. Further, while the problem could in principle be seen as a component of a general predator-prey analysis (Elliott, et al., 2006); as presented, the focus is on setting up the physics. The biological implications are not discussed. (For more discussion of this problem, see (McNeill Alexander, 2006).) We modified the problem to retain the physics tasks, but corrected the biology.

### *Growing a worm*

While the cheetah-pronghorn example is a useful physics task in a biological context,[12] it represents too small a part of the predator-prey interaction to have much biological significance. In the next example, we discuss a negotiation to create a physics problem with stronger biological authenticity.

One of our most interesting negotiations occurred early in our creation of the NEXUS physics class. We begin the class with a topic that is essential to developing a good physics perspective: dimensional analysis and scaling (functional dependence). Although scientists in many disciplines use this method, physicists consider it crucial to their worldview. Mathematical quantities in physics are viewed not just as numbers, but as structured quantities that are defined by how they transform when arbitrary decisions used in describing the system (coordinate system, measurement scale, etc.) are changed. (Redish, 2005)

We began our discussion with our physicist asking our biologist for a biologically-relevant problem on scaling and functional dependence. Our biologist responded that he had a problem that he used in his principle-based organismal biology class. This is shown in figure 3.

> A typical specimen of the common earthworm (*Lumbricus terrestris*) exhibits the following mean dimensions: Weight - 3.7 g, Length – 12 cm, Width – 0.64 cm; Oxygen consumption of the body - 0.98 µmole $O_2$/g; Oxygen absorption across the skin - 2.4 nmole $O_2$/mm$^2$ h
>
> 1) It is realistic to model the shape of the earthworm as a solid cylinder. Using the dimensions of a typical earthworm above, calculate its surface area (ignore the surface areas of the blunt ends in all calculations), volume, and density.
>
> 2) Earthworms are aerobic organisms that require oxygen to carry out cellular respiration in order to make ATP. They absorb oxygen from the surrounding soil across their moist skin, which means that the rate of oxygen absorption depends on their surface area. Almost all of the cells in their bodies consume oxygen in cellular respiration, which mean that the rate of oxygen consumption depends on their mass, which is directly proportional to their volume.
>
>    - Calculate the ratio of surface area to volume (S/V ratio) of this typical earthworm.
>    - Predict how a decrease in this ratio might have a profound effect on the respiratory activity (i.e., oxygen consumption) of an earthworm.
>
> 3) If you assume that an earthworm were to absorb and consume oxygen at the typical rates above, then show whether or not it can absorb sufficient oxygen to maintain the measured respiratory rate of its entire body.
>
> 4) In theory, organisms can grow via two ways
>
> 4A) An organism can simply grow by increasing its length without changing its other dimensions. Assume that our typical earthworm has doubled its length with no other changes in its dimensions.
>
>    - Calculate new S/V ratio of this enlarged earthworm
>    - Calculate whether oxygen absorption rate can keep pace with its oxygen consumption rate
>
> 4B) However, many organisms grow isometrically, meaning that each linear dimension increases by the same factor.
>
>    - Assuming that our typical earthworm has doubled its length, what factor will its surface area increase by?
>    - What factor will its volume increase by?
>    - Calculate the new S/V ratio of this enlarged earthworm.
>    - Calculate whether oxygen absorption rate can keep pace with its oxygen consumption rate
>    - Predict what will happen to this enlarged organism.
>
> 5) Tropical earthworms can reach sizes up to 4 m. Predict how these large worms manage to balance their oxygen absorption rate against their respiratory rate.

*Fig. 3: The biologist's form of the scaling problem*

Our physicist thought this problem was an appropriate starting point, but thought it would be useful to add two parts to the problem: using graphs and deriving an explicit equation. These are shown in figure 4.

These physics-style tasks underline the focus in physics classes on learning to use general-purpose tools and multiple representation translation, while the original biology problem focuses on specific calculations and the implications for the organism.

> A1) Consider the effect of changing the various size parameters of a worm. First consider a worm of length 12 cm that grows by keeping its length the same but increasing its radius. Use a spreadsheet to plot the total oxygen absorbed through the skin of the worm and the total oxygen used by the worm as a function of its length from a radius of 0 cm (not really reasonable) up to a radius of 1 cm. Do the two curves cross? Explain what the crossing means and what its implications are.
>
> A2) Now consider a worm width 0.64 cm that grows by keeping its width the same but increasing its length. Use a spreadsheet to plot the total oxygen absorbed through the skin of the worm and the total oxygen used by the worm as a function of its length from a length of 0 cm (not really reasonable) up to a length of 50 cm. Do the two curves cross? Explain what the crossing means and what its implications are.
>
> B) Consider a general cylindrical organism of density $d$ (g/cm$^3$), length $h$ (cm) and radius $R$ (cm). If the rate of oxygen absorption through the skin is $A$ (mole/cm$^2$-h) and the rate at which is uses oxygen is $B$ (mole/g-h), write a symbolic expression for the total rate of oxygen absorbed by the worm and the total rate of oxygen used by the worm. Find the maximum radius the worm could be before it would have a problem taking in enough oxygen.

*Fig. 4: The physicists' additions to the biologists' scaling problem.*

The second part of the physics problem is particularly revealing. The skin area of a cylindrical organism of radius $R$ and length $h$ is $2\pi Rh$ (ignoring the end caps) and its volume is $\pi R^2 h$. This gives its mass as $m = \pi R^2 h d$. As a result, the general condition that says the skin must be able to absorb oxygen at a rate faster than it is used in the volume can be written

$$A(2\pi Rh) > B(\pi R^2 hd)$$

We notice that the $h$ cancels out telling us that the length of the worm is not the constraining parameter. Solving for $R$ yields the condition

$$R < \frac{2A}{Bd}.$$

This shows that there is a maximum radius for a cylindrical worm for the given parameters.

When our physicist presented this result to our working team of mixed physicists and biologists, the reaction was striking. To a person, the physicists' reaction was, "Oooh! Way cool!" Whereas the biologists' reaction was, "Well, yeah, but that's not very interesting because that's not the way organisms grow." The physicists felt they had learned something of value from the equation; the biologists had not.

After considerable negotiation, our final result accepted both ideas: that for a physics class, the development of representational translation skills (expressing results as numbers, graphs, and equations) was of value. But the problem was reworked to emphasize the biological realities of the situation. Furthermore, we came up with a version that demonstrated biological authentic value: the implications of the scaling analysis for explaining the value of certain variations and the implications for a deeper understanding of phylogenetic development (of gills and lungs). The final version also made more explicit the modeling assumptions that go into the analysis, something both disciplines felt was valuable. The final result is shown in figure 5.

> The earthworm absorbs oxygen directly through its skin. The worm does have a good circulatory system (with multiple small hearts) that brings the oxygen to all the cells. But the cells are distributed through the worm's volume and the oxygen only gets to come in through the skin -- so the surface to volume ratio plays an important role. Let's see how this works. Here are the worm's parameters.
>
> A typical specimen of the common earthworm (*Lumbricus terrestris*) has the following average dimensions: Mass - 3.7 g, Length – 12 cm, Width – 0.64 cm.
>
> The skin of the worm can absorb oxygen at a rate of $A = 0.24$ μmole (μmole = $10^{-6}$ moles) per square cm per hour. The body of the worm needs to use approximately $B = 0.98$ μmole (μmole = $10^{-6}$ moles) of oxygen per gram of worm per hour.
>
> A. It is reasonable to model the shape of the earthworm as a solid cylinder? Using the dimensions of a typical earthworm above, calculate its surface area (ignore the surface areas of the blunt ends in all calculations), volume, and density.
>
> B. If the worm is much longer than it is wide ($L >> R$) is it OK to ignore the end caps of the cylinder in calculating the surface area? How does the surface area and volume of the worm depend on the length of the worm, $L$, and the radius of the worm, $R$?
>
> C. For an arbitrary worm of length $L$, radius $R$, and density $d$, write an equation (using the symbols $A$ and or $B$ rather than the numbers) that expresses the number of moles of oxygen the worm absorbs per hour and the number of moles the worm uses per hour. What is the condition that the worm takes in oxygen at a rate fast enough to survive? Does this simple model predict that the typical worm described above absorbs sufficient oxygen to survive?
>
> D.1. Consider the effect of changing the various size parameters of a worm. First consider a worm of length 12 cm that grows by keeping its length the same but increasing its radius. Use a spreadsheet to plot the total oxygen absorbed through the skin of the worm and the total oxygen used by the worm as a function of its length from a radius of 0 cm (not really reasonable) up to a radius of 1 cm. Do the two curves cross? Explain what the crossing means and what its implications are.
>
> D.2. Now consider a worm width 0.64 cm that grows by keeping its width the same but increasing its length. Use a spreadsheet to plot the total oxygen absorbed through the skin of the worm and the total oxygen used by the worm as a function of its length from a length of 0 cm (not really reasonable) up to a length of 50 cm. Do the two curves cross? Explain what crossing means and what its implications are.
>
> D.3. Write (in symbols) an equation that represents the crossover condition -- that the oxygen taken in per hour exactly equals the oxygen used per hour. Cancel common factors. Discuss how this equation tells you about what you learned about worm growth by doing the two graphs.
>
> E. Our analysis in D was a modeling analysis. An organism like an earthworm might grow in two ways: by just getting longer or isometrically -- by scaling up all its dimensions. What can you say about the growth of an earthworm by these two methods as a result of your analysis in part D? Does a worm have a maximum size? If so, in what sense? If so, find it.
>
> F. In typical analyses of evolution and phylogenetic histories, earthworm-like organisms are the ancestors of much larger organisms than the limit here permits. Discuss what sort of variations in the structure of an earthworm might lead to an organism that solves the problem of growing isometrically larger than the limit provided by this simple model.

*Fig. 5: The negotiated compromise scaling problem*

## V. Inferences: How the disciplinary worldviews differ

These are just a few examples of the many times we saw physicists and biologists demonstrating distinct perspectives on the nature of the knowledge they see as appropriate to present in introductory classes. In some cases, the distinct perspectives are deeper than simply, "What is appro-priate content to present in an introductory class?" Sometimes they represent epistemological differences in the nature of the way scientists perceive the knowledge in their disciplines.

Since both biology and physics are extremely diverse professions, we are not claiming the differences we have observed in any way distinguish "being a biologist" from "being a physicist." Rather we see these as "cultural averages" that are common and create strong pressures on instructional methods, particularly at an introductory level.

Some of the cultural differences we have observed are *epistemological* – they represent different ways that instructors view the nature of the knowledge in their discipline. Others are *pedagogical* – they represent cultural differences in what instructors (especially at the introductory level) view as appropriate to do in their classes.

Since we have not done a quantitative sociological study, but are simply codifying our observations of hundreds of hours of personal discussions, we refer to our conclusions as *heuristics* – experience-based guidelines for proceeding in the absence of firm laws or principles.

**Disciplinary epistemological sticking points**

Here are some of the differences that we often find lead to "interesting discussions" between us. These are important points to be aware of and good places to begin your interdisciplinary conversations.

### *Physics: Common cultural components*

- Introductory physics classes often stress *reasoning from a few fundamental (usually mathematically formulated) principles.*
- Physicists often stress building a complete understanding of the *simplest possible (often highly abstract) examples* – "toy models" – and often don't go beyond them at the introductory level.
- Physicists *quantify* their view of the physical world, *model with math*, and *think with equations*, qualitatively as well as quantitatively.
- Physicists concern themselves with *constraints* that hold no matter what the internal details (conservation laws, center of mass, ...).

These elements will be familiar to anyone who has ever taught introductory physics. What is striking is that we usually do not articulate this for students – and <u>none</u> of these elements are typically present in an introductory biology class. Biologists have other concerns.

### *Biology: Common cultural components*

- Biology is often *complex*. Many biological processes involve the interactions of component parts leading to emergent phenomena, which includes the property of life itself.
- Most introductory *biology does not emphasize quantitative reasoning* and problem solving to the extent that it serves in introductory physics.
- Biology contains a critical *historical constraint* in that natural selection can only act on pre-existing molecules, cells, and organisms for generating new solutions.
- Much of introductory biology is *descriptive* (and introduces a large vocabulary)
- However, biology – even at the introductory level – looks for *mechanism* and often considers micro-macro connections between the molecules involved and the larger phenomenon.
- Biologists (both professionals and students) focus on and value *real examples* and *structure-function relationships*.

We note that while there is overlap in some aspects of our introductory science classes, they tend to be treated differently. For example, some of the biological models used in introductory classes can be described as toy models – highly unrealistic and introduced for the purpose of understanding one component of a mechanism. The Hardy-Weinberg model of evolution is one such, relying on unrealistic assumptions.

However, some of our biologists considered traditional toy-model physics examples, such as the simple harmonic oscillator (mass-on-a-spring), irrelevant, uninteresting, and useless until the physicists were able to show its value as a starting-point model for many real-world and relevant biological examples. This required making it clear from the first that a Hooke's law oscillator was an *oversimplified model* and illustrating how it would be modified for realistic cases. This is unfortunately rarely done in introductory physics classes and physicists are typically taken to task for "wanting to live in frictionless vacuums."

The specific examples we give in this article illustrate the implication of some of these differences in how physicists and biologists view instructional issues. As we see in the discussion of the fluid flow and worm problems, biologists, while using equations, often focus on the calculational value of the equations for particular realistic or semi-realistic cases. The physicists use the equations to extract abstract insights whether or not any direct applications are obvious.

All the examples discussed in this paper illustrate the high value that biologists place on functional implications, something the physicists tend to ignore. Both groups are interested in having their students learn "mechanistic reasoning" but interpret it in different ways when applied in introductory classes. Biologists often tend to focus on molecular chemical level mechanisms and to insist on the connection to some functionality; physicists (in courses at the undergraduate level) tend to focus on macroscopic structures and how "what is happening" is controlled by general mathematical principles.

**Disciplinary pedagogical heuristics**

In addition to the epistemological differences each discipline brought to our discussions, it also became clear that there were strong cultural differences in the pedagogy that the two disciplines were accustomed to bringing to bear in introductory classes. One key difference is in the role played by homework and problem solving.

Physics as a profession broadly sees problem solving as the key to the discipline. Even in high school and introductory college classes and even in the most traditional and unreformed physics classes, problem-solving plays a central role. Every introductory textbook has dozens, sometimes hundreds of problems provided at the end of each chapter. In some contexts, this may devolve into a selection of simple mathematical exercises ("plug-and-chug" problems), but few physics instructors would see that as desirable or appropriate.

The goal of physics teachers is to have their students able to apply the broad general principles they are learning to new situations and examples. While in part this is due to the powerful role that mathematical reasoning plays even in introductory physics, it has the implication that even the most traditional physics class has a strong active learning component.

Introductory classes in biology often have little or no homework or problem solving (with the exception of units on genetics). What we have learned from our example of the principle-based organismal biology class is that trying to teach principle-based reasoning *without* the aid of problem-solving homework can be extremely difficult.

Perhaps because of this long tradition of problem solving in physics, modern physics pedagogy is changing to emphasize increased active engagement of students within the classroom. As a result of what is now a long history of physics education research (PER), many active learning envi-ronments are spreading through the physics community – and many research results are becoming available supporting the value of these environments. (McDermott & Redish, 1999; Meltzer & Thornton, 2012) While this is perhaps a historical artifact of physics having developed a strong educational research community, it provides another place where we can learn from each other and begin to try to understand what has been learned from PER that can transfer to biology and what biology education research has to reinvent in a way that is true to its discipline.

## Conclusion: Learning enough to respect – and challenge

From our extended conversations, both with each other and with other biologists, chemists, and physicists, we conclude that, "science is not just science." Scientists in each discipline employ a tool kit of different types of scientific reasoning. A particular discipline is not characterized by the exclusive use of a set of particular reasoning types, but each discipline is characterized by the tendency to emphasize some types more than others and to value different kinds of knowledge differently. The physicist's enthusiasm for characterizing an object as a disembodied point mass can make a biologist uncomfortable because they often find in biology that function is directly related to structure. Yet, similar sorts of simplified structures can be very powerful in some biological analyses.

The enthusiasm that some biologists feel toward our students learning physics is based not so much on the potential for students to learn physics knowledge, but rather on the potential for them to learn the types of reasoning more often experienced in physics classes. They *don't* want their students to think like physicists. They want them to think like biologists who have access to many of the tools and skills physicists introduce in introductory physics classes.

Indeed, the recognition of biology as a multidisciplinary field, as expressed in the *Vision and Change* report (AAAS, 2011), is based on a collective sense that other disciplines more readily employ different types of reasoning, and one of our goals as educators of biology students is to help them gain the facility to apply appropriate types of scientific reasoning for addressing different biological problems.

The exercise of characterizing the tendency toward different disciplines emphasizing different reasoning is an important step toward more effective teaching of scientific reasoning in all classes taken by biology students. And one step toward that is to have the disciplines understand the different perspectives and values they each bring to their science.

In order to include physics in biology classes and develop physics classes with authentic value for biologists, biologists don't have to become physicists and physicists don't have to become biologists. But each group needs to develop an understanding of the other's discipline; not just the content but also their epistemological style and goals.

While we have detailed our experience, each interdisciplinary exploration is bound to be unique to the individuals and institutions involved. But what we expect will be common to such activities will include: showing respect for each others' discipline and insights, a willingness to reconsider one's own discipline from a different point of view; and finally, patience, persistence, and humor.

We conclude that the process is significantly more complex than many reformers working largely within their discipline often assume. But the process of learning each others' ropes – at least to the extent that we can understand each other's goals and ask each other challenging questions – can be both enlightening and enjoyable. And much to our surprise, we each feel that we have developed a deeper understanding of our own discipline as a result of our discussions.

**ACKNOWLEDGMENTS**

We would like to thank the University of Maryland Physics Education Research Group, Biology Education Research Group, and the NEXUS Physics team for the many discussions that have informed the views expressed here. The authors would also like to thank Chris Bauer, Melanie Cooper, Catherine Crouch, Mike Klymkowsky, and Jessica Watkins for their discussions on issues of interdisciplinarity. This work is supported NSF-TUES 09-19816, NSF-TUES DUE 11-22818, and the Howard Hughes Medical Institute NEXUS grant.

[1] While there have been numerous attempts to create courses in "Physics for the Life Sciences" or "Physics for Pre-Meds" over the past few decades, none of them have gained much traction for reasons we will discuss below.

[2] We note the extended collaborations of Bialek & Botstein at Princeton and Meredith & Bolker at UNH.

[3] UMd BERG: [http://umdberg.pbworks.com/w/page/8039417/FrontPage]

[4] UMd PERG: [http://www.physics.umd.edu/perg/]

[5] NSF CCLI 09-19816, *The Physics of Life: Interdisciplinary Education at the Introductory Level*

[6] NSF DRL 00-87519, *Learning How to Learn Science: Metacognition in post-secondary physics education for bioscience majors*; NSF DUE 03-41447, *Helping Students Learn How to Learn: Open-source physics worksheets integrated with TA development resources*; NSF DRL 04-40113, *Toward a new Conceptualization of What Constitutes Progress in Learning Physics, K-16: Resources, frames, and networks*; NSF DUE 07-15567, *Collaborative Research: Open-source physics tutorial worksheets with faculty/TA development and implementation resources.*

[7] In actual fact, the fluid has a *velocity profile*, the fluid at the walls moving with 0 velocity and the fluid in the center of the pipe moving the fastest. A complete analysis requires vector calculus, but our toy model gives a correct result – if the proportionality constants needed are taken from the more advanced analysis.

[8] Physics version of the artery problem: [http://www.physics.umd.edu/perg/abp/TPProbs/Problems/M/M22.htm]

[9] Note that our biologist and physicist each insist on using different notations for the quantities represented in the H-P equation. This problem is widespread throughout the curriculum as each community has distinct cultural standards and notational practices.

[10] Project NEXUS (HHMI) [http://www.hhmi.org/grants/office/nexus/]

[11] Project NEXUS (UMCP) [http://umdberg.pbworks.com/w/page/44091483/Project NEXUS UMCP]

[12] The modified problem is available at: [http://umdberg.pbworks.com/w/page/44332396/The%20cat%20and%20the%20antelope]